\documentclass[aps,prl,reprint,groupedaddress]{revtex4-1}

\usepackage{color}
\usepackage{xcolor}
\usepackage{float}
\usepackage{graphicx}
\usepackage{subfigure}
\usepackage{mathtools}

\definecolor{goodgreen}{RGB}{0,128,0}

\begin{document}


\title{New Electromagnetic Modes in Space-Time Modulated Dispersion-Engineered Media}


\author{Nima Chamanara}
\author{Zo\'{e}-Lise Deck-L\'{e}ger }
\author{Christophe Caloz}
\affiliation{Polytechnique Montr\'{e}al, Montr\'{e}al, Qu\'{e}bec H3T 1J4, Canada.}

\author{Dikshitulu Kalluri}
\affiliation{University of Massachusetts Lowell, MA USA.}


\date{\today}

\begin{abstract}
We report on new electromagnetic modes in space-time modulated dispersion-engineered media. These modes exhibit unusual dispersion relation, field profile and scattering properties. They are generated by coupled codirectional space-time harmonic pairs, and occur in space-time periodic media whose constituent materials exhibit specific dispersion. Excitation of a slab of such a medium with \emph{subluminal} modulation results in periodic transfer of energy between the incident frequency and a frequency shifted by a multiple of the modulation frequency, whereas \emph{superluminal} modulation generates exponentially growing frequencies. These modes may find applications in optical mixers, terahertz sources and other optical devices.
\end{abstract}


\maketitle


Periodic structures are an essential part of modern photonic and microwave technologies. Electromagnetic bandgaps emerging from such structures play a crucial role in many applications, including photonic crystal waveguides~\cite{joannopoulos2011photonic}, filters~\cite{collin2007foundations}, fiber gratings~\cite{kashyap2009fiber}, etc. These bandgaps support evanescent electromagnetic modes characterized by complex propagation numbers i.e. exponential decay in space~\cite{joannopoulos2011photonic}.

The temporal counterparts~\cite{kalluri2016electromagnetics} of conventional photonic crystals, termed \emph{temporal photonic crystals}, have been shown to exhibit complex frequencies or vertical bandgaps, as opposed to horizontal bandgaps in spatial photonic crystals~\cite{zurita2010resonances, zurita2009reflection}. They represent electromagnetic media whose constitutive parameters vary periodically in time~\cite{zurita2010resonances, zurita2009reflection}. Their vertical bandgaps describe instabilities~\cite{reyes2015observation, cassedy1967dispersion}, where the amplitude of electromagnetic waves grow/decay exponentially, i.e. with $e^{\pm\omega t}$ time dependence, everywhere in space.

Combining \emph{both} time and space modulations allows one to control not only the orientation -- horizontal or vertical -- of the bandgaps but also their oblique alignment~\cite{cassedy1963dispersion, chamanara2016asymmgap, taravati2017nonreciprocal}. \emph{Subluminal} space-time modulated structures, where the phase velocity of the modulation ($v_\text{m}$) is smaller than the velocity of light in the background medium ($v_\text{b}$), support obliquely aligned horizontally oriented bandgaps, whereas \emph{superluminal} space-time modulated with $v_\text{m}>v_\text{b}$ support obliquely aligned vertically oriented bandgaps~\cite{taravati2017nonreciprocal}.

This paper considers periodic space-time modulated media composed of \emph{dispersive} materials. It shows that the addition of dispersion to space-time modulation brings about a unique diversity of electromagnetic modes. Particularly, such a medium can support new modes with dispersion, field and scattering properties that are radically different from those accessible in conventional or space-time periodic media composed of \emph{nondisperive} materials.

\begin{figure}[ht!]
\centering
\subfigure{ \label{fig:conc-ND-crossing}
\includegraphics[width=0.47\columnwidth, page=1]{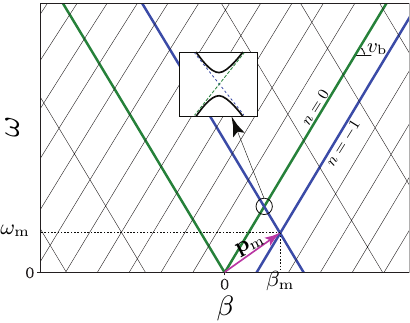}}
\subfigure{ \label{fig:conc-DE-crossing}
\includegraphics[width=0.47\columnwidth, page=2]{pics.pdf}}
\caption{Comparison of dispersion diagrams of space-time modulated periodic media composed of nondipersive and dispersive background materials for $M\rightarrow 0$. (a)~Case of nondispersive background material. (b)~Case strongly dispersive background material. The magenta arrow shows the modulation (or space-time period) vector ($\mathbf{p}_\text{m}$). Here, unessentially, the modulation for both cases subluminal ($v_\text{m}=\omega_\text{m}/\beta_\text{m}<v_\text{b}$). For clarity, the two space-time harmonics $n=0$ and $n=-1$ are emphasized.}
\label{fig:conc-crossing}
\end{figure}

Consider first the case space-time periodic media composed of nondisperive materials. The corresponding permittivity may be written as
\begin{equation} \label{eq:mod-eps-nondisp-f}
\epsilon(\mathbf{r}, t)=\epsilon_0\epsilon_\text{r}\left[1+M f_\text{per}\left(\omega_\text{m} t-\beta_\text{m}z\right)\right],
\end{equation}
where $\epsilon_\text{r}$ is the permittivity of the nondispersive background medium, $M$ is the modulation depth and $f_\text{per}(.)$ is a periodic function with period $2\pi$. The dispersion relation of the nondispersive background medium ($\epsilon_\text{r}$) is plotted in green in Fig.~\ref{fig:conc-ND-crossing}. Electromagnetic waves in the modulated medium can be expressed in the Floquet-Bloch form~\cite{cassedy1963dispersion, chamanara2016asymmgap, taravati2017nonreciprocal}
\begin{equation} \label{eq:bloch-E}
\mathbf{E}(\mathbf{r}, t)=e^{j\left(\omega t-\beta z\right)}\sum\limits_{n=-\infty}^{\infty}\mathbf{E}_{n}e^{jn\left(\omega_\text{m} t-\beta_\text{m}z\right)}.
\end{equation}
It can be shown from the Fourier expansion in~\eqref{eq:bloch-E} that dispersion curves are periodic along the vector $\mathbf{p}_\text{m}=\beta_\text{m} \hat{\mathbf{\beta}} + \omega_\text{m} \hat{\mathbf{\omega}}$~~\cite{cassedy1963dispersion}. Typical dispersion curves for a vanishingly small modulation depth ($M\rightarrow 0$) are plotted in Fig.~\ref{fig:conc-ND-crossing}. Since the green lines are solutions to dispersion relations in this limit, the rest of the dispersion diagram is obtained by periodically shifting the background dispersion line along $\mathbf{p}_\text{m}$. This provides an accurate estimation of the dispersion curves everywhere except at the points where these shifted curves, or space-time harmonics, intersect. The two space-time harmonics forming such intersections are phase matched and hence strongly couple to one another. For a nondispersive background medium, such couplings involve a forward harmonic and a backward harmonic, as emphasized by the circle in Fig.~\ref{fig:conc-ND-crossing}. In the case of subluminal modulation, as in Fig.~\ref{fig:conc-ND-crossing}, the resulting forward-backward coupling generates evanescent modes, with complex $\beta$ corresponding to horizontally oriented gaps, as shown in the inset. For a superluminal modulation, the forward-backward coupling between space-time harmonics generates unstable modes, with complex $\omega$ corresponding to vertically oriented gaps (inset of Fig.~\ref{fig:conc-ND-crossing} rotated by 90$^\circ$)~\cite{cassedy1963dispersion, taravati2017nonreciprocal}.

In such a space-time modulated periodic medium, composed of \emph{nondispersive} materials, only forward-backward coupling between space-time harmonics is possible, as illustrated in the inset of Fig.~\ref{fig:conc-crossing}. In contrast, if the background medium is dispersion-engineered, as in Fig.~\ref{fig:conc-DE-crossing}, then it is possible to produce codirectional, forward-forward (emphasized by the circle) or backward-backward, coupling harmonic pairs. This paper investigates such scenarios in a space-time periodic dispersion-engineered medium with Lorentz background material dispersion, and the resulting unusual electromagnetic modes emerging from such codirectional couplings for both subluminal and superluminal modulations.

The background material in the dispersive-background medium is described by the following relative Lorentz permittivity
\begin{equation}
\epsilon_\text{r}\left(\omega\right) = 1 + \frac{\omega_p^2}{\omega_\text{r}^2 + j\gamma\omega - \omega^2},
\end{equation}
where $\omega_\text{r}$ is the Lorentz resonance frequency,  $\omega_p = e\sqrt{\frac{n_a}{\epsilon_0 m_e}}$ is the plasma frequency, $\gamma$ is the relaxation rate and $n_a$ is the molecular (atomic) density of the material. We assume that the material density is spatio-temporally modulated as
\begin{equation} \label{eq:n-rt}
n_a(\mathbf{r}, t)=n_{0}\left[1+M f_\text{per}\left(\omega_\text{m} t-\beta_\text{m}z\right)\right],
\end{equation}
which may be practically achieved by a modulated laser in a nonlinear material or an acoustic wave in an acousto-optic material.

To find the electromagnetic modes in this medium, equations
\begin{equation} \label{eq:wave-st}
\nabla\times\nabla\times\mathbf{E}\left(\mathbf{r}, t\right)+\mu\frac{\partial^{2}}{\partial t^{2}}\mathbf{D}\left(\mathbf{r}, t\right)=0,
\end{equation}
\begin{equation} \label{eq:D-P}
\mathbf{D}\left(\mathbf{r}, t\right) = \epsilon_{0}\mathbf{E}\left(\mathbf{r}, t\right)+\mathbf{P}\left(\mathbf{r}, t\right),
\end{equation}
\begin{equation} \label{eq:P-eom}
\frac{\partial^{2}}{\partial t^{2}}\mathbf{P}\left(\mathbf{r}, t\right)+\gamma\frac{\partial}{\partial t}\mathbf{P}\left(\mathbf{r}, t\right)+\omega_{r}^{2}\mathbf{P}\left(\mathbf{r}, t\right)=\frac{n_a\left(\mathbf{r}, t\right)e^{2}}{m_e}\mathbf{E}\left(\mathbf{r}, t\right),
\end{equation}
which are the wave equation, the displacement field equation and the classical Newton equation for a Lorentz medium~\cite{jackson2007classical}, respectively, are solved self-consistently. The solution of the problem is found by expressing all the field quantities in these equations in terms of the space-time Floquet-Bloch expansion
\begin{equation} \label{eq:Bloch-E}
\mathbf{\Psi}(\mathbf{r}, t)=e^{j\left(\omega t-\beta z\right)}\sum\limits_{n=-\infty}^{\infty}\mathbf{\Psi}_{n}e^{jn\left(\omega_\text{m} t-\beta_\text{m}z\right)},
\end{equation}
\noindent where $\mathbf{\Psi} = \mathbf{E}, \mathbf{P}, \mathbf{D}$, and $\mathbf{\Psi}_n = \mathbf{E}_n, \mathbf{P}_n, \mathbf{D}_n$ are complex constants, and the space-time dispersive curves may be written as $[\omega_n,\beta_n]=[\omega+n\omega_\text{m},\beta+n\beta_\text{m}]$~\cite{supplementalMaterial}.

\begin{figure}[ht!]
\centering
\subfigure{
\includegraphics[width=1.0\columnwidth, page=3]{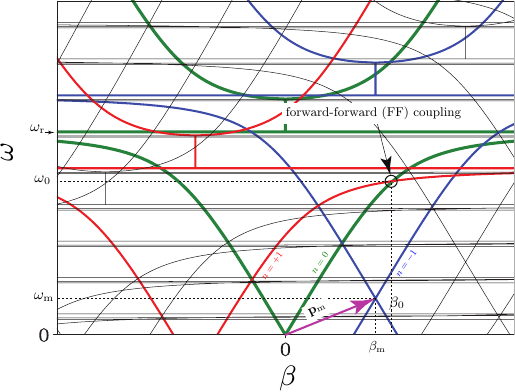}}
\caption{Dispersion diagram of a subluminally space-time modulated lossless Lorentz medium for $M\rightarrow 0$. The magenta arrow represents the modulation vector. The three space-time harmonics $n=0$ (green), $n=+1$ (red) and $n=-1$ (blue) are emphasized.}
\label{fig:up-conv-sub-periodic-modes}
\end{figure}

Consider, as an example, a lossless Lorentz medium with normalized parameters $\omega_p/\omega_\text{r} = 0.6$ and $\gamma/\omega_\text{r}\rightarrow 0$ (lossless). The background material density is modulated as in \eqref{eq:n-rt}, with spatial and temporal modulation frequencies $\beta_\text{m} = 0.8673 \omega_\text{r}/c$ and $\omega_\text{m} = 0.18 \omega_\text{r}$, respectively. The corresponding dispersion curves are plotted in Fig.~\ref{fig:up-conv-sub-periodic-modes} for an $M\rightarrow 0$, where the crossing (or phase matched) space-time harmonics $n=0$ and $n=+1$ is emphasized by a circle.
%

\begin{figure}[ht!]
\centering
\subfigure{
\includegraphics[width=0.8\columnwidth, page=4]{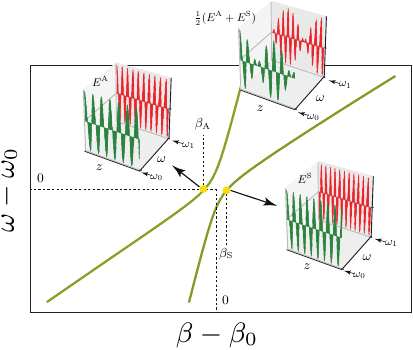}}
\caption{Dispersion diagram at the crossing circled in Fig.~\ref{fig:up-conv-sub-periodic-modes} for the finite modulation depth $M=0.01$. The solid curves represent the real parts of the dispersion, while the insets show the field profiles for the additive ($E^\text{A}$) and subtractive ($E^\text{S}$) modes, as well as their superposition ($\frac{1}{2}(E^\text{A}+E^\text{S})$). The imaginary parts are zero.}
\label{fig:crossing-sub-real}
\end{figure}

In the case of a finite nonzero modulation depth, these space-time harmonics couple to each other as a result of phase matching. The dispersion relation and electromagnetic fields of the modes emerging from such forward-forward (FF) coupling are computed in~\cite{supplementalMaterial} as
\begin{align}	\label{eq:modes-sym-antisym}
E_A = a_0 e^{j(\omega_0 t-\beta_A z)} + a_1 e^{j[(\omega_0+N\omega_\text{m}) t - (\beta_A+N\beta_\text{m}) z]}, \\
E_S = a_0 e^{j(\omega_0 t-\beta_S z)} - a_1 e^{j[(\omega_0+N\omega_\text{m}) t - (\beta_S+N\beta_\text{m}) z]},
\end{align}
where the subscripts 'A' and 'S' stand for ``additive'' and ``subtractive'', respectively, in reference with the signs in~\eqref{eq:modes-sym-antisym}, and are plotted in Fig.~\ref{fig:crossing-sub-real} for $M=0.01$. These modes have purely real, but slightly distinct, wavenumbers, and contain the frequencies of the coupled space-time harmonics at the crossing point, i.e. $e^{j\omega_0 t}$ for their green part and $e^{j(\omega_0+\omega_\text{m})t}$ for their red part [Eq.~\eqref{eq:Bloch-E}]. Therefore, their temporal spectrum consists of two frequencies, $\omega_0$ and $\omega_1=\omega_0+N\omega_\text{m}$, where $N$ is the index difference between the crossing space-time harmonics ($N=1$ in Fig.~\ref{fig:up-conv-sub-periodic-modes}). The consequence of such dispersion relations and field profiles will be explained later when electromagnetic scattering from such media will be discussed.

\begin{figure}[ht!]
\centering
\subfigure{
\includegraphics[width=1.0\columnwidth, page=5]{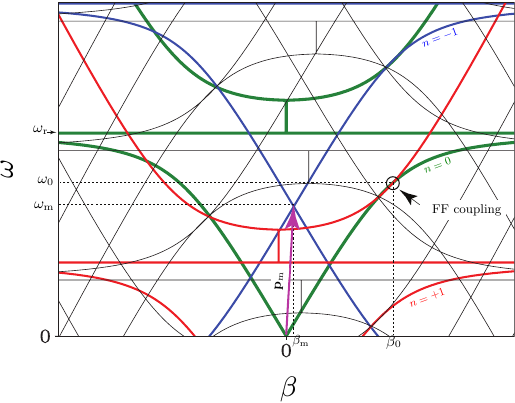}}
\caption{Dispersion curves of a superluminal space-time modulated Lorentz medium for $M\rightarrow 0$. The magenta arrow represents the modulation vector.}
\label{fig:up-conv-super-periodic-modes}
\end{figure}

Next consider a superluminal modulation for a lossless Lorentz medium with the same background parameters as in the subluminal case ($\omega_p/\omega_\text{r} = 0.6$ and $\gamma/\omega_\text{r}=0$). The density of the medium is again modulated as in \eqref{eq:n-rt} but with the superluminal spatial-temporal modulation $[\beta_\text{m},\omega_\text{m}]= [0.0724 \omega_\text{r}/c,0.64 \omega_\text{r}]$. The corresponding dispersion curves are plotted in Fig.~\ref{fig:up-conv-super-periodic-modes} for $M\rightarrow 0$, where the crossing (or phase matched) space-time harmonics $n=0$ and $n=+1$ is emphasized by a circle.
%
%
In the case of a finite modulation depth, these space-time harmonics couple to each other as a result of phase matching. The dispersion relation and electromagnetic fields of the modes emerging from such forward-forward (FF) coupling are computed in~\cite{supplementalMaterial} as
\begin{align}	\label{eq:modes-grow-decay}
E_G = a_0 e^{j(\omega_0 t-\beta_G z)} + a_1 e^{j[(\omega_0+N\omega_\text{m}) t - (\beta_G+N\beta_\text{m}) z]}, \\
E_D = a_0 e^{j(\omega_0 t-\beta_D z)} - a_1 e^{j[(\omega_0+N\omega_\text{m}) t - (\beta_D+N\beta_\text{m}) z]},
\end{align}
where the subscripts 'G' and 'D' stand for ``growing'' and ``decaying'', respectively, and are plotted in Fig.~\ref{fig:crossing-sub-real} for $M=0.01$. These modes have complex conjugate wavenumbers, $\beta_G = \beta + j\alpha$ and $\beta_D = \beta - j\alpha$, corresponding to exponentially growing and decaying modes along $z$, respectively. As in the subluminal case, their temporal spectrum consists of the two frequencies $\omega_0$ and $\omega_0+N\omega_\text{m}$ where $N$ is the index difference between the crossing space-time harmonics. In contrast to evanescent modes in conventional (purely spatial) photonic crystals, which also form a complex conjugate pair, these modes carry real power with Poynting vector positive for both $E_G$ and $E_D$, i.e. both modes correspond to forward propagation (along $+z$) and can be excited by a copropagating incident wave.
\begin{figure}[ht!]
\centering
\subfigure{
\includegraphics[width=0.8\columnwidth, page=6]{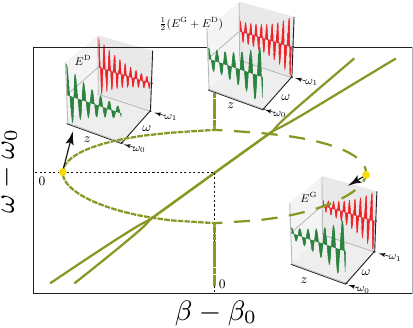}}
\caption{Dispersion diagram at the crossing circled in Fig.~\ref{fig:up-conv-super-periodic-modes} for a finite modulation depth $M=0.01$. The solid curves represent the real parts. The dashed curves represent the imaginary parts and the insets show the field profiles for growing ($E^\text{G}$) and decaying ($E^\text{D}$) fields, as well as their superposition ($\frac{1}{2}(E^\text{G}+E^\text{D})$).}
\label{fig:crossing-super-real-imag}
\end{figure}

Next a slab of a space-time periodic Lorentz material is excited by a plane wave, and reflected and transmitted fields as well as the modes excited inside the slab are computed using the full-wave mode matching technique~\cite{supplementalMaterial}. This technique provides a great deal of physical insight into the physics of the system since it provides exact information on the contribution of the unbounded slab contributing to the different waves. Moreover, it is immune of most of the numerical errors that plague other full-wave numerical techniques, such as numerical dispersion error, and errors associated with absorbing boundary conditions and excitation models.

In the subluminal case, the incident wave must be matched to the additive and subtractive modes~\eqref{eq:modes-sym-antisym} in order to excite them effectively. Note that at $z=0$, the superposition of the two modes, $E = \frac{1}{2}(E_A + E_S) = a e^{j\omega_0 t}$, include only the frequency $\omega_0$ and, since $\beta_A \approx \beta_S \approx \beta_0$, an incident plane wave $E^\text{i} = \hat{\mathbf{x}} e^{j(\omega_0 t - \beta_0 z)}$ is well matched to $E = \frac{1}{2}(E_A + E_S)$ and excites $E_A$ and $E_S$ equally. Therefore, the incident medium should either have the permittivity $\epsilon_{r1}=(\omega_0/c\beta_0)^2$ or be matched to such a permittivity through a matching section, in order to excite these modes efficiently.

\begin{figure}[ht!]
\centering
\includegraphics[width=0.9\columnwidth, page=7]{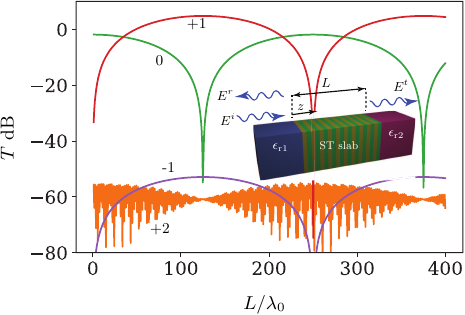}
\caption{Transmission, $T=|\mathbf{E}^\text{t}|/|\mathbf{E}^\text{i}|$, through a slab of the medium in Fig.~\ref{fig:crossing-sub-real} excited by a plane wave versus its length $L$. The numbers $n$ represent the amplitude of the transmitted harmonic at $\omega_0 + n\omega_\text{m}$.}
\label{fig:slab-up-T-sub-z}
\end{figure}

Figure~\ref{fig:slab-up-T-sub-z} shows the spectrum of the field transmitted through the slab versus its length, for a slab with subluminal space-time modulated Lorentz medium in Fig.~\ref{fig:crossing-sub-real} excited by a plane wave $E^i = \hat{\mathbf{x}} e^{j(\omega_0 t - \beta_0 z)}$. Note that electromagnetic energy is periodically transferred between the harmonics at $\omega_0$ and $\omega_0+\omega_\text{m}$, while the rest of the harmonics are more than $50$~dB weaker. At the coherence length, a maximum of energy, possibly with gain due the energy of the modulation, is transferred to the $\omega_0+\omega_\text{m}$ harmonic, without any undesirable intermodulation effects, in contrast to what occurs in conventional modulators or mixers. This process is similar to energy exchange between two forward coupled waveguides. Since the additive and subtractive modes in~\eqref{eq:modes-sym-antisym} have a slightly different propagation constant, their harmonics at $\omega_0$ acquire some gradual phase mismatch as they propagate along the structure, until they completely fall out of phase at the coherence length, while their harmonics at $\omega_0+\omega_\text{m}$ arrive in phase at the coherence length, and the process is next reversed, and repeated periodically. Note that similar transitions may occur between different states of a space-time modulated photonic crystal, termed \emph{interband photonic transitions} in the photonics literature in analogy to electronics~\cite{winn1999interband, yu2009complete}. In this sense, the transitions in Fig.~\ref{fig:slab-up-T-sub-z} may be called \emph{intraband} photonic transitions.

\begin{figure}[ht!]
\centering
\includegraphics[width=0.9\columnwidth, page=8]{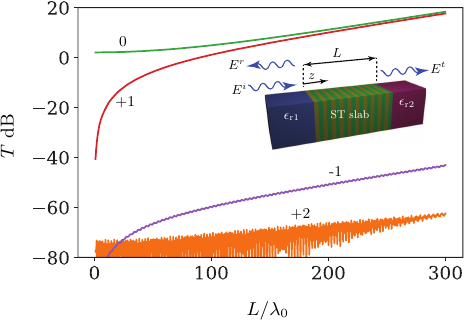}
\caption{Transmission, $T=|\mathbf{E}^\text{t}|/|\mathbf{E}^\text{i}|$, from a slab of the medium in Fig.~\ref{fig:crossing-super-real-imag} through by a plane wave, versus its length $L$. The numbers $n$ represent the amplitude of the transmitted harmonic at $\omega_0 + n\omega_\text{m}$.}
\label{fig:slab-up-T-super-z}
\end{figure}

As in the subluminal case, in the superluminal space-time modulated dispersion engineered medium shown in Fig.~\ref{fig:crossing-super-real-imag}, an incident wave at frequency $\omega_0$ excites both the growing and decaying modes. Figure~\ref{fig:slab-up-T-super-z} presents scattering from a slab of the superluminal space-time modulated medium in Fig.~\ref{fig:crossing-super-real-imag}. Note that the transmitted harmonics at $\omega_0$ and $\omega_0+\omega_\text{m}$ grow exponentially as the length of the slab is increased, while the rest of the harmonics are more than $60$~dB weaker. The growing mode $E_G$ dominates the exponentially decaying mode $E_D$ as the length of the slab is increased, and since the growing mode contains both harmonics at $\omega_0$ and $\omega_0+\omega_\text{m}$, these harmonics grow exponentially and are transmitted when they reach the other end of the slab.

The reported unusual electromagnetic modes may find applications in efficient harmonic generators and perfect mixers as they provide unprecedented control over the generation of electromagnetic harmonics without undesired intermodulation effects. Moreover, since the generated harmonics are exponentially amplified, such modes can be used to produce electromagnetic sources in frequency ranges that are not easily accessible, such as terahertz gap.

\bibliography{ReferenceList2_abbr}

\begin{thebibliography}{15}%
\makeatletter
\providecommand \@ifxundefined [1]{%
 \@ifx{#1\undefined}
}%
\providecommand \@ifnum [1]{%
 \ifnum #1\expandafter \@firstoftwo
 \else \expandafter \@secondoftwo
 \fi
}%
\providecommand \@ifx [1]{%
 \ifx #1\expandafter \@firstoftwo
 \else \expandafter \@secondoftwo
 \fi
}%
\providecommand \natexlab [1]{#1}%
\providecommand \enquote  [1]{``#1''}%
\providecommand \bibnamefont  [1]{#1}%
\providecommand \bibfnamefont [1]{#1}%
\providecommand \citenamefont [1]{#1}%
\providecommand \href@noop [0]{\@secondoftwo}%
\providecommand \href [0]{\begingroup \@sanitize@url \@href}%
\providecommand \@href[1]{\@@startlink{#1}\@@href}%
\providecommand \@@href[1]{\endgroup#1\@@endlink}%
\providecommand \@sanitize@url [0]{\catcode `\\12\catcode `\$12\catcode
  `\&12\catcode `\#12\catcode `\^12\catcode `\_12\catcode `\%12\relax}%
\providecommand \@@startlink[1]{}%
\providecommand \@@endlink[0]{}%
\providecommand \url  [0]{\begingroup\@sanitize@url \@url }%
\providecommand \@url [1]{\endgroup\@href {#1}{\urlprefix }}%
\providecommand \urlprefix  [0]{URL }%
\providecommand \Eprint [0]{\href }%
\providecommand \doibase [0]{http://dx.doi.org/}%
\providecommand \selectlanguage [0]{\@gobble}%
\providecommand \bibinfo  [0]{\@secondoftwo}%
\providecommand \bibfield  [0]{\@secondoftwo}%
\providecommand \translation [1]{[#1]}%
\providecommand \BibitemOpen [0]{}%
\providecommand \bibitemStop [0]{}%
\providecommand \bibitemNoStop [0]{.\EOS\space}%
\providecommand \EOS [0]{\spacefactor3000\relax}%
\providecommand \BibitemShut  [1]{\csname bibitem#1\endcsname}%
\let\auto@bib@innerbib\@empty
\bibitem [{\citenamefont {Joannopoulos}\ \emph {et~al.}(2011)\citenamefont
  {Joannopoulos}, \citenamefont {Johnson}, \citenamefont {Winn},\ and\
  \citenamefont {Meade}}]{joannopoulos2011photonic}%
  \BibitemOpen
  \bibfield  {author} {\bibinfo {author} {\bibfnamefont {J.~D.}\ \bibnamefont
  {Joannopoulos}}, \bibinfo {author} {\bibfnamefont {S.~G.}\ \bibnamefont
  {Johnson}}, \bibinfo {author} {\bibfnamefont {J.~N.}\ \bibnamefont {Winn}}, \
  and\ \bibinfo {author} {\bibfnamefont {R.~D.}\ \bibnamefont {Meade}},\
  }\href@noop {} {\emph {\bibinfo {title} {Photonic crystals: molding the flow
  of light}}}\ (\bibinfo  {publisher} {Princeton university press},\ \bibinfo
  {year} {2011})\BibitemShut {NoStop}%
\bibitem [{\citenamefont {Collin}(2007)}]{collin2007foundations}%
  \BibitemOpen
  \bibfield  {author} {\bibinfo {author} {\bibfnamefont {R.~E.}\ \bibnamefont
  {Collin}},\ }\href@noop {} {\emph {\bibinfo {title} {Foundations for
  microwave engineering}}}\ (\bibinfo  {publisher} {John Wiley \& Sons},\
  \bibinfo {year} {2007})\BibitemShut {NoStop}%
\bibitem [{\citenamefont {Kashyap}(2009)}]{kashyap2009fiber}%
  \BibitemOpen
  \bibfield  {author} {\bibinfo {author} {\bibfnamefont {R.}~\bibnamefont
  {Kashyap}},\ }\href@noop {} {\emph {\bibinfo {title} {Fiber {B}ragg
  gratings}}}\ (\bibinfo  {publisher} {Academic press},\ \bibinfo {year}
  {2009})\BibitemShut {NoStop}%
\bibitem [{\citenamefont {Kalluri}(2010)}]{kalluri2016electromagnetics}%
  \BibitemOpen
  \bibfield  {author} {\bibinfo {author} {\bibfnamefont {D.~K.}\ \bibnamefont
  {Kalluri}},\ }\href@noop {} {\emph {\bibinfo {title} {Electromagnetics of
  time varying complex media: frequency and polarization transformer}}}\
  (\bibinfo  {publisher} {CRC Press},\ \bibinfo {year} {2010})\BibitemShut
  {NoStop}%
\bibitem [{\citenamefont {Zurita-S{\'a}nchez}\ and\ \citenamefont
  {Halevi}(2010)}]{zurita2010resonances}%
  \BibitemOpen
  \bibfield  {author} {\bibinfo {author} {\bibfnamefont {J.~R.}\ \bibnamefont
  {Zurita-S{\'a}nchez}}\ and\ \bibinfo {author} {\bibfnamefont
  {P.}~\bibnamefont {Halevi}},\ }\href@noop {} {\bibfield  {journal} {\bibinfo
  {journal} {Phys. Rev. A}\ }\textbf {\bibinfo {volume} {81}},\ \bibinfo
  {pages} {053834} (\bibinfo {year} {2010})}\BibitemShut {NoStop}%
\bibitem [{\citenamefont {Zurita-S{\'a}nchez}\ \emph
  {et~al.}(2009)\citenamefont {Zurita-S{\'a}nchez}, \citenamefont {Halevi},\
  and\ \citenamefont {Cervantes-Gonzalez}}]{zurita2009reflection}%
  \BibitemOpen
  \bibfield  {author} {\bibinfo {author} {\bibfnamefont {J.~R.}\ \bibnamefont
  {Zurita-S{\'a}nchez}}, \bibinfo {author} {\bibfnamefont {P.}~\bibnamefont
  {Halevi}}, \ and\ \bibinfo {author} {\bibfnamefont {J.~C.}\ \bibnamefont
  {Cervantes-Gonzalez}},\ }\href@noop {} {\bibfield  {journal} {\bibinfo
  {journal} {Phys. Rev. A}\ }\textbf {\bibinfo {volume} {79}},\ \bibinfo
  {pages} {053821} (\bibinfo {year} {2009})}\BibitemShut {NoStop}%
\bibitem [{\citenamefont {Reyes-Ayona}\ and\ \citenamefont
  {Halevi}(2015)}]{reyes2015observation}%
  \BibitemOpen
  \bibfield  {author} {\bibinfo {author} {\bibfnamefont {J.}~\bibnamefont
  {Reyes-Ayona}}\ and\ \bibinfo {author} {\bibfnamefont {P.}~\bibnamefont
  {Halevi}},\ }\href@noop {} {\bibfield  {journal} {\bibinfo  {journal} {Appl.
  Phys. Lett.}\ }\textbf {\bibinfo {volume} {107}},\ \bibinfo {pages} {074101}
  (\bibinfo {year} {2015})}\BibitemShut {NoStop}%
\bibitem [{\citenamefont {Cassedy}(1967)}]{cassedy1967dispersion}%
  \BibitemOpen
  \bibfield  {author} {\bibinfo {author} {\bibfnamefont {E.}~\bibnamefont
  {Cassedy}},\ }\href@noop {} {\bibfield  {journal} {\bibinfo  {journal} {Proc.
  IEEE}\ }\textbf {\bibinfo {volume} {55}},\ \bibinfo {pages} {1154} (\bibinfo
  {year} {1967})}\BibitemShut {NoStop}%
\bibitem [{\citenamefont {Cassedy}\ and\ \citenamefont
  {Oliner}(1963)}]{cassedy1963dispersion}%
  \BibitemOpen
  \bibfield  {author} {\bibinfo {author} {\bibfnamefont {E.}~\bibnamefont
  {Cassedy}}\ and\ \bibinfo {author} {\bibfnamefont {A.}~\bibnamefont
  {Oliner}},\ }\href@noop {} {\bibfield  {journal} {\bibinfo  {journal} {Proc.
  IEEE}\ }\textbf {\bibinfo {volume} {51}},\ \bibinfo {pages} {1342} (\bibinfo
  {year} {1963})}\BibitemShut {NoStop}%
\bibitem [{\citenamefont {Chamanara}\ \emph {et~al.}(2017)\citenamefont
  {Chamanara}, \citenamefont {Taravati}, \citenamefont {Deck-L\'eger},\ and\
  \citenamefont {Caloz}}]{chamanara2016asymmgap}%
  \BibitemOpen
  \bibfield  {author} {\bibinfo {author} {\bibfnamefont {N.}~\bibnamefont
  {Chamanara}}, \bibinfo {author} {\bibfnamefont {S.}~\bibnamefont {Taravati}},
  \bibinfo {author} {\bibfnamefont {Z.-L.}\ \bibnamefont {Deck-L\'eger}}, \
  and\ \bibinfo {author} {\bibfnamefont {C.}~\bibnamefont {Caloz}},\ }\href
  {\doibase 10.1103/PhysRevB.96.155409} {\bibfield  {journal} {\bibinfo
  {journal} {Phys. Rev. B}\ }\textbf {\bibinfo {volume} {96}},\ \bibinfo
  {pages} {155409} (\bibinfo {year} {2017})}\BibitemShut {NoStop}%
\bibitem [{\citenamefont {Taravati}\ \emph {et~al.}(2017)\citenamefont
  {Taravati}, \citenamefont {Chamanara},\ and\ \citenamefont
  {Caloz}}]{taravati2017nonreciprocal}%
  \BibitemOpen
  \bibfield  {author} {\bibinfo {author} {\bibfnamefont {S.}~\bibnamefont
  {Taravati}}, \bibinfo {author} {\bibfnamefont {N.}~\bibnamefont {Chamanara}},
  \ and\ \bibinfo {author} {\bibfnamefont {C.}~\bibnamefont {Caloz}},\
  }\href@noop {} {\bibfield  {journal} {\bibinfo  {journal} {arXiv Prepr.
  arXiv:1705.06311}\ } (\bibinfo {year} {2017})}\BibitemShut {NoStop}%
\bibitem [{\citenamefont {Jackson}(2007)}]{jackson2007classical}%
  \BibitemOpen
  \bibfield  {author} {\bibinfo {author} {\bibfnamefont {J.~D.}\ \bibnamefont
  {Jackson}},\ }\href@noop {} {\emph {\bibinfo {title} {Classical
  electrodynamics}}}\ (\bibinfo  {publisher} {John Wiley \& Sons},\ \bibinfo
  {year} {2007})\BibitemShut {NoStop}%
\bibitem [{sup()}]{supplementalMaterial}%
  \BibitemOpen
  \href@noop {} {\bibinfo  {journal} {See the supplemental materials at [URL]
  for detailed analysis of the space-time modulated slab}\ }\BibitemShut
  {NoStop}%
\bibitem [{\citenamefont {Winn}\ \emph {et~al.}(1999)\citenamefont {Winn},
  \citenamefont {Fan}, \citenamefont {Joannopoulos},\ and\ \citenamefont
  {Ippen}}]{winn1999interband}%
  \BibitemOpen
\bibfield  {journal} {  }\bibfield  {author} {\bibinfo {author} {\bibfnamefont
  {J.~N.}\ \bibnamefont {Winn}}, \bibinfo {author} {\bibfnamefont
  {S.}~\bibnamefont {Fan}}, \bibinfo {author} {\bibfnamefont {J.~D.}\
  \bibnamefont {Joannopoulos}}, \ and\ \bibinfo {author} {\bibfnamefont
  {E.~P.}\ \bibnamefont {Ippen}},\ }\href@noop {} {\bibfield  {journal}
  {\bibinfo  {journal} {Phys. Rev. B}\ }\textbf {\bibinfo {volume} {59}},\
  \bibinfo {pages} {1551} (\bibinfo {year} {1999})}\BibitemShut {NoStop}%
\bibitem [{\citenamefont {Yu}\ and\ \citenamefont
  {Fan}(2009)}]{yu2009complete}%
  \BibitemOpen
  \bibfield  {author} {\bibinfo {author} {\bibfnamefont {Z.}~\bibnamefont
  {Yu}}\ and\ \bibinfo {author} {\bibfnamefont {S.}~\bibnamefont {Fan}},\
  }\href@noop {} {\bibfield  {journal} {\bibinfo  {journal} {Nat. Photon.}\
  }\textbf {\bibinfo {volume} {3}},\ \bibinfo {pages} {91} (\bibinfo {year}
  {2009})}\BibitemShut {NoStop}%
\end{thebibliography}%

\end{document}